\documentclass[12pt]{article}

\usepackage{scicite}
\usepackage{graphicx}
\usepackage{times}
\usepackage{amsmath}

\newenvironment{sciabstract}{%
\begin{quote} \bf}
{\end{quote}}

\newcounter{lastnote}

\title{Observation of Non-Vanishing Optical Helicity in Thermal Radiation from Symmetry-Broken Metasurfaces}

\author
{Xueji Wang,$^{1\bot}$ Tyler Sentz,$^{1\bot}$ Sathwik Bharadwaj,$^{1}$ Subir Kumar Ray,$^{1}$ \\ Yifan Wang,$^{1}$ Dan Jiao,$^{1}$ Limei Qi,$^{2}$ Zubin Jacob$^{1*}$\\
\\
\normalsize{$^{1}$Elmore Family School of Electrical and Computer Engineering, Purdue University}\\
\normalsize{West Lafayette, Indiana, 47907, USA}\\
\normalsize{$^{2}$School of Electronic Engineering, Beijing University of Posts and Telecommunications}\\ \normalsize{Beijing, 100876, China}\\
\\
\normalsize{$^{\bot}$These authors contribute equally to this work.}
\\
\normalsize{$^{\ast}$To whom correspondence should be addressed; E-mail:  zjacob@purdue.edu.}
}

\date{}

\begin{document} 


\baselineskip24pt


\maketitle


\begin{sciabstract}
Abstract: Spinning thermal radiation is a unique phenomenon observed in condensed astronomical objects including the Wolf–Rayet star EZ-CMa and the red degenerate star G99-47, due to existence of strong magnetic fields. Here, by designing symmetry-broken metasurfaces, we demonstrate that spinning thermal radiation with a non-vanishing optical helicity can be realized even without applying a magnetic field. We design non-vanishing optical helicity by engineering a dispersionless band which emits omnidirectional spinning thermal radiation, where our design reaches 39\% of the fundamental limit. Our results firmly suggest metasurfaces can impart spin coherence in the incoherent radiation excited by thermal fluctuations. The symmetry-based design strategy also provides a general pathway for controlling thermal radiation in its temporal and spin coherence.
\end{sciabstract}


\section*{Introduction}

Thermal radiation describes the universal phenomenon that all objects at non-zero temperatures emit infrared electromagnetic energy\cite{Review1,Review_NRM,Fan_Review}. Notable research progress has been made so far in tailoring its temporal coherence, i.e. spectrum\cite{Greffet1D,MM_TEGold1Antenna,MM_SemiSirods}, and the spatial coherence, i.e. directivity \cite{Direction2,Direction1,MP_Focusing}. However, the photon spin, another crucial characteristic of electromagnetic radiation, is commonly ignored, since most thermal emitters show weak to zero spin angular momentum (SAM) in the emitted waves. Surprisingly, the thermal radiation reaching the earth from many astronomical objects possesses substantial circular polarization. The unique phenomenon provides strong evidence for the presence of a magnetic field around stars\cite{RedDegen,EZ_CMa,Star2}, or reveals the existence of chiral organic molecules\cite{Star1}.  Recent studies also show that circular polarization can be a possible sign of life\cite{Star3}. Therefore, revealing photon spin characteristics in thermal radiation is of fundamental interest as it contains unique information regarding the emitters. 

In analogy with celestial observations, it has recently come to light that the application of an external magnetic field or intrinsic non-reciprocity in  generic bianisotropic media also leads to circularly polarized thermal emission characteristics. This is true for magneto-optical materials\cite{CP_MO}, topological Weyl semimetals\cite{WCP,Axion}, and topological insulators\cite{TopologicalInsulator1,TopologicalInsulator2}. This phenomenon of spinning thermal radiation is yet to be observed due to  material constraints and the requirement of magnetic fields. When time-reversal symmetry is preserved such as in hyperbolic materials\cite{SKL}, the circularly polarized thermal radiation is predicted at specific angles\cite{SKL} but the overall purity is very small\cite{TopologicalInsulator2}. Also, in the hyperbolic case, the thermo-optic helicity, which is the summation of the projected SAM over the far-field hemisphere, is always zero/near-zero. We note that recent studies have experimentally demonstrated spinning thermal radiation using mechanisms such as the optical Rashba effect \cite{shitrit2013spin} and integrated waveplate \cite{Boreman}. However, the spinning thermal radiation is limited to the surface-normal direction with integrated waveplate\cite{Boreman}. In the optical Rashba effect, the spinning thermal radiation is only generated in oblique directions away from the surface normal\cite{shitrit2013spin}. The mirror symmetry guarantees that the LCP and RCP appear in pairs, which causes the net optical helicity to be zero. 

In this paper, we show that photon spin introduces an additional degree of freedom for tailoring thermal radiation (Fig.~1, A to D). Even though spectrally and angularly tailored thermal radiative sources based on plasmon/phonon polaritons\cite{ZMZ_SPP,Mark_Dynamic,Qiangli_SPhP}, bound states in the continuum\cite{Alu_BIC}, epsilon-near-zero resonances\cite{ENZ_Body}, hyperbolic materials\cite{Yuri_MHB,ENZ_MMExp}, photonic crystals\cite{PC_2DCore} and quantum wells\cite{PC_NP} have received extensive attention, an omnidirectional circularly polarized light source is still not accessible, especially in the mid-infrared region.  Our thermal metasurface approach is compact so large benchtop light sources and bulky optical components (waveplates and polarizers) are not required.

We note that extensive efforts have been made so far to generate spin-selective absorption from microwave to visible frequencies \cite{R1_Detection,R2_Mirror,R3_NIR,R4_Coupling,R5_Chiral_F,R7_Visible,R8_APLMirror,R9_BIC,R10_DualBand,R11_Thin,R6_PMirror}. Our work provides a direct experimental demonstration proving that spin-selectivity can be realized in the radiation excited by incoherent thermal fluctuations. Additionally, the efforts on spin-selectivity are mainly limited to the surface-normal direction. In contrast, the non-vanishing optical helicity demonstrated here extends the effort to control the angular distribution of photon spin with symmetry-based designs. Although recent numerical studies have proposed omnidirectional spin-selective absorption\cite{R6_PMirror}, the experimental realization is obstructed due to the complicated three-dimensional architecture. Our 2D planar design is compatible with the well-established nano/micro-fabrication techniques, making it practical for future energy applications.

Our non-magnetic approach is summarized in Fig.~1, E to G. For conventional materials when both inversion- and mirror- symmetries are preserved, photon spin of thermal photons is degenerate in energy and momentum space and zero circularly polarized heat radiation is generated (Fig.~1E). Breaking inversion symmetry can lead to spin characteristics in oblique directions away from the surface normal, as shown in the extrinsic chirality \cite{ExtChi_APL, ExtChi_LSA} and the optical Rashba effect \cite{shitrit2013spin}. However, the anti-symmetric spin pattern causes the net helicity to vanish once again (Fig.~1F). To overcome these limitations, in this work, we show that the simultaneous breaking of mirror symmetry and inversion symmetry releases all the constraints. Remarkably, we obtain an asymmetric spin pattern and non-vanishing optical helicity in the radiation excited by thermal fluctuations (Fig.~1G). We point out that non-vanishing optical helicity only requires the breaking of mirror symmetry. The simultaneous breaking of mirror and inversion symmetries is required to obtain asymmetric pattern of spinning thermal radiation. The complete symmetry breaking in our design also provides freedom to introduce symmetries. We demonstrate three examples of symmetry based thermal engineering  with mirror, inversion, and four-fold rotational (C4) symmetries, and show that incoherent thermal fluctuations can be substantially altered by these symmetries. With this symmetry-based approach, we also suggest a general and effective pathway to engineering the spectral and spin characteristics of thermal radiation.

\section*{Results}

To generate non-vanishing optical helicity, we start by designing a symmetry-broken metasurface that lifts spin degeneracy. A schematic of the metasurface is shown in Fig.~2A, where a rectangular array of F-shape meta-atoms is patterned on a silicon dioxide (SiO$_2$) dielectric layer with a gold (Au) backplane. Our `F' meta-atoms are designed to break both inversion- and mirror- symmetry in this quasi-2D system. Our designed structure also provides a large parameter space for optimizing the spectrum and energy-momentum-spin band structure of the thermal emission. As shown in Fig.~2B, the measured emissivity of left-handed circular-polarized heat radiation (LCP) is more than 4 times larger than the emissivity of right-handed circular-polarized heat radiation (RCP) at the surface normal direction. In Fig.~2C, we show a complete characterization of the polarization state of thermal radiation based on Stokes parameters, where $S_3/S_0$ indicates the purity of the circular polarization. $S_3/S_0$ is equivalent to the widely adopted metric `degree of circular polarization (DoCP)', which is expressed as $(\sigma_+ - \sigma_-)/(\sigma_+ + \sigma_-)$, and $\sigma_+$ and $\sigma_-$ are the emissivities of LCP and RCP, respectively. We represent the radiation state emanating from incoherent thermal fluctuations in the metasurface by a circular area that is centered close to the north pole of the Poincar\'e sphere (inset of Fig.~2C). We emphasize that the near-zero $S_1$ and $S_2$ at 7 $\mu$m indicate the relatively limited DoCP is actually from the limited coherence of the thermal emission signal. In Fig.~2D, we plot $S_0$ and degree of polarization (DoP) to demonstrate the total intensity and the polarization purity of the thermal radiation from our metasurface.  

To show the microscopic mechanisms of the circular polarization and non-vanishing optical helicity, we investigate the near-field electromagnetic response of the metasurface. The major mechanisms of spin degeneracy removal in photonic systems can be divided into two categories. On the one hand, the Pancharatnam-Berry phase generated by a geometric gradient can lead to spin-dependent photon behaviors \cite{SpinHall1,Uriel_Gradient,ErezHasman_Gradient,PB_Chiral}. On the other hand, the phenomenon can also originate from the intrinsic local chirality in the meta-atoms \cite{CP1,CP2, Chiral_3D}. We reveal that the strong intrinsic local chirality of our F-shape plasmonic meta-atoms is the origin of the observed circular-polarized thermal radiation in the far field. To show this, we simulate the electric field strength at 7$\mu$m under LCP and RCP excitations (Fig.~2E). We calculate the local absorption cross-section density $\alpha_{abs}^{(l)}\left(\mathbf{r}^{\prime}, \omega\right)=\operatorname{Im}\left[\varepsilon\left(\mathbf{r}^{\prime}, \omega\right)\right]\left[\omega\left|\mathbf{E}^{(l)}\left(\mathbf{r}^{\prime}, \omega\right)\right|^2\right]/\left[c|\mathbf{E}_{inc}^{(l)}\left(\omega\right)|^2\right]$, where $\varepsilon\left(\mathbf{r}^{\prime}, \omega\right)$ is the  permittivity at point $\mathbf{r}^{\prime}$ and frequency $\omega$,  $\mathbf{E}^{(l)}\left(\mathbf{r}^{\prime}, \omega\right)$ is the $(l)$-polarized local electric field, and $\mathbf{E}_{inc}^{(l)}\left(\omega\right)$ is the incident plane wave. According to the local Kirchhoff Law \cite{NonEqu_Greffet}, the local emissivity density is equal to the local absorption cross-section density. We map the local emissivity density of LCP and RCP radiation in the meta-atom in Fig.~2F. An evident resonant enhancement can be observed for LCP radiation near the region of the two horizontal lines of ‘F’, while the enhancement is clearly absent for RCP. The strong spin-selective local dissipation/emission in the meta-atoms is the direct reason for the strong LCP emission peak at 7µm. We note that the F-shape meta-atoms can also be seen as perturbed resonators (see Supplementary Materials Figure~S19). The symmetry breaking introduced by the perturbative segments lifts the spin degeneracy in the resonators and thus results in spinning thermal radiation. The perturbation also provides a ‘knob’ to effectively tune the temporal and spin coherence of thermal radiation \cite{Alu_BIC,R9_BIC}.

To demonstrate the non-vanishing optical helicity and map the energy-momentum ($E-k$) dispersion of thermal radiation, we establish a unique spin-polarized angle-resolved thermal emission spectroscopy (SPARTES) system. The system allows us to characterize the spectral and polarimetric properties of thermal radiation signals at various deflection ($\theta$) and azimuth angles ($\phi$) over the far-field hemisphere as shown in Fig.~3A. (The detailed description of the SPARTES system can be found in Materials and Methods. Additional thermal emission spectroscopy data can be found in Supplementary Materials.) We now consider the experimental data of thermal photons with momentum along the x-axis i.e. along $k_x$ for our discussion ($k_y=0$). The measured spectra are plotted in Fig.~3, H to K and show an excellent agreement with simulations (Fig.~3, D to G). A strong contrast between the emissivity of LCP and RCP is observed, which is a direct manifestation of the spin degeneracy removal. 

First, we note that the LCP and RCP emissivities ($\sigma_+,  \sigma_-$) are not only distinct (Fig.~3, D, E, H, I) but also asymmetric in momentum space ($+k$ and $-k$). However, there is a clear symmetric pattern in energy-momentum space of the averaged emissivity defined as $\sigma_{avg}=\sigma_+/2 + \sigma_-/2$ (Fig.~3, F and J). This is because the averaged emissivity is proportional to the total radiative power, and the symmetric total radiation in k-space is fundamentally guaranteed by the reciprocity in this magnetic-field-free system. The asymmetric LCP and RCP emissivity also leads to a unique DoCP pattern, which is neither symmetric nor anti-symmetric. We point out that this spin asymmetry in k-space is from the complete symmetry breaking of our ‘F’ structure, where neither the 2D inversion-symmetry nor the mirror-symmetry is preserved.

To achieve net non-zero helicity, it is necessary to obtain spin asymmetry for a large solid angle beyond the normal emission. We demonstrate this strong effect at 7µm by engineering a dispersionless energy band along $k_x$. To reveal the thermal radiation characteristics in this dispersionless energy band, we plot the differential circular-polarized emissivity ($\sigma_+ - \sigma_-$) at 7µm on the far-field hemisphere (projected onto the $k_{xy}$ plane). As evidently illustrated in Fig.~3B and Fig.~3C, the thermal radiation is predominantly left-handed circularly polarized in all directions. The total optical helicity $H$ is nonzero and is defined as\cite{Helicity},
\begin{equation}
H=\sum_{k} \hbar\left(n_{k, \sigma+}-n_{k, \sigma-}\right)  \propto  \sum_{k} \left(\sigma_+ - \sigma_-\right)
\end{equation}
where $n_{k,\sigma+}$ and $n_{k,\sigma-}$ are the photon numbers of LCP and RCP associated with the wavevector $k$.  We note that total SAM associated with the radiation can be characterized by optical helicity and its vectorial counterpart optical spin\cite{Helicity}. However, optical helicity is more well-suited for describing the efficiency of imparting SAM through far-field thermal radiation (see Supplementary Materials Figure~S15). 

To explore the fundamental limit, we normalize the optical helicity $H$ of our device by the optical helicity $H_0$ of a perfect omnidirectional LCP thermal emitter (where the differential emissivity $\sigma_{0+} - \sigma_{0-} = 1$). We show our design reaches 39\% of the fundamental limit at 7µm  from the experimental results, i.e. $\left.{H}/{H_0}\right|_{\lambda=7\mu m}=\left.{\sum_k\left(\sigma_{+}-\sigma_{-}\right)}/{\sum_k\left(\sigma_{0+}-\sigma_{0-}\right)}\right|_{\lambda=7 \mu m}=39 \%$. This is in strong contrast with previous demonstrations where the optical helicity is identically  zero\cite{shitrit2013spin} since the spin characteristics are anti-symmetric in k-space ($n_{k,\sigma+} = n_{-k,\sigma-}$).

Now, we show that the spin characteristics of thermal photons can be tailored by the symmetries of the metasurface. We design the ‘F’ structure to have complete symmetry breaking i.e. none of the mirror symmetry, inversion symmetry, or high-order rotational symmetries are present in the meta-atoms themselves. Therefore, by introducing additional symmetries at a larger length scale than the meta-atoms, we demonstrate effective control of thermal radiation in its spectral and spin properties. To show this, in Fig.~4, we present the $E-k$ dispersion where thermal photon momentum is along the y-axis ($k_y \neq 0, k_x=0$). The experimentally measured averaged emissivity and the corresponding DoCP in three different metasurfaces with mirror, inversion, and 4-fold rotational symmetries (C4) are shown. We point out three interesting features in the obtained thermal emission spectra. In Fig.~4C, by introducing mirror-symmetry along the real-space y-axis, the DoCP is made to vanish. In the second row, we show that the DoCP of thermal radiation in the inversion-symmetric device becomes symmetric in energy-momentum space (Fig.~4F), which is in strong contrast to the ‘single-F’ metasurface where the DoCP is asymmetric (Fig.~3 G and K). Thirdly, we design a C4 device which also possesses inversion symmetry (equivalent to C2 in this 2D case).  Here, distinctive spectral, spatial, and spin features are present as a manifestation of the additional rotation symmetry in this C4 symmetric thermal metasurface (Fig.~4G) as compared to the C2 symmetric structure (Fig.~4D). We note that the near-zero DoCP originates from the coupling between the four ‘F’ meta-atoms in a unit cell and can be explained through a geometric phase mechanism\cite{PBPhase} (See Supplementary Materials Figure~S7). We emphasize that all averaged emissivity plots (Fig.~4, B, E and H) show symmetric patterns in k-space irrespective of spatial symmetries, which arises from the requirement of reciprocity.

\section*{Discussion}

Finally, we discuss the impact and potential applications of our results. The observed non-vanishing optical helicity indicates highly omnidirectional and even broadband circularly polarized thermal emitters can be realized through metasurfaces. Our demonstrated devices here can be directly applied as a wide-angle, narrow-band circular-polarized mid-infrared light source, for applications including optical gas sensing\cite{GasSensing} and infrared imaging\cite{InfaredImaging}. We highlight that the metasurface-based device can be integrated into on-chip systems to achieve compact functional devices. Moreover, considering the field of near-field heat transfer, we expect a thermal emitter with non-vanishing optical helicity will have the unique ability to transfer net angular momentum and torque through heat\cite{CP_MO, TopologicalInsulator1}.  The exploration in this direction may facilitate the discovery of thermal spin torque phenomena at the nanoscale, and eventually, benefit the development of thermal energy devices including near-field spin thermophotovoltaics. Additionally, the unique spectral-spatial-spin feature of the engineered thermal emission can be exploited as high-contrast infrared beacons in outdoor environments, as the background thermal emission from other natural objects is highly incoherent without any spin textures. Finally, our metasurface-based strategy for controlling optical spin introduces an additional degree of freedom of thermal radiation engineering, providing a general approach to tailoring thermal radiation for future energy applications.

\section*{Materials and Methods}
\subsection*{Device fabrication}
The devices shown in this work consist of rectangular arrays of gold meta-atoms patterned on thin films of silicon dioxide (SiO$_2$). To fabricate the device, we first deposited a 180 nm gold (Au) layer on a silicon (Si) wafer. The Au layer prevents the thermal radiation of the heater and the Si substrate from being measured. Then 820nm SiO$_2$ layer was deposited on the Au layer. We note that 10nm titanium (Ti) layers were also used at the Si/Au and the Au/SiO$_2$ interfaces to increase adhesion. The F-shape meta-atoms were first written to the spin-coated PMMA 950 A4 resist using a 100 keV electron beam lithography system (JEOL JBX-8100FS). After the resist was developed, 120nm Au was deposited on the patterned resist using an e-beam evaporator. Finally, the metasurface was created in the lift-off process, where the excess resist was removed with acetone. The area of the fabricated metasurface was 5mm x 5mm, relaxing the alignment requirements for angle-resolved measurements.

\subsection*{Spin-polarized angle-resolved thermal emission spectroscopy (SPARTES)}
We used a 2-axis rotary sample stage in the SPARTES setup to characterize the thermal emission along different directions over the far-field hemisphere (see Supplementary Materials Figure~S3 and Figure~S4). For sample heating, a 88.9mm ring heater controlled through a programmable power supply was used. A thin tungsten sheet was attached to the ring heater to cut off the thermal emission from the heater. Tungsten was chosen here because of its low emissivity, good thermal conductivity, and high-temperature stability. The thermal radiation signal from a sample was first collected and collimated through a parabolic mirror. Then a waveplate and a linear polarizer were employed to extract the spin properties of the thermal radiation. A Fourier transform infrared spectrometer (FTIR, Thermo Fisher Scientific is50 Nicolet) with a liquid-nitrogen-cooled mercury-cadmium-telluride (MCT) detector was used for spectral analysis. The entire system was in a nitrogen-purged environment to eliminate the atmospheric absorption of the mid-infrared thermal radiation signal.

\subsection*{Polarimetric analysis}

To characterize the Stokes parameters of thermal radiation signal, we used a retardation waveplate followed by a linear polarizer (see Supplementary Materials Figure S5). Assuming the transmission axis of the polarizer is at an angle $\alpha$ to the x-axis; the fast axis of the waveplate is at an angle $\beta$ to the x-axis; and the phase difference between the fast and slow axis of the waveplate is $\delta$; the transmitted light intensity for incident radiation with Stokes parameters $(S_0, S_1, S_2, S_3)$ is \cite{Stokes1, Stokes2}
\begin{equation}
\begin{aligned}
&I\left(\alpha, \beta\right)=\frac{1}{2}\left\{S_0\right. \\
&\quad+S_1\left[\cos 2 \beta \cos 2\left(\alpha-\beta\right) - \cos  \delta \sin 2 \beta \sin 2\left(\alpha-\beta\right)\right] \\
&\quad+S_2\left[\sin 2 \beta \cos 2\left(\alpha-\beta\right) + \cos \delta \cos 2 \beta \sin 2\left(\alpha-\beta\right)\right] \\
&\quad+S_3 \sin \delta \sin 2\left(\alpha-\beta\right)\}
\end{aligned}
\end{equation}

The Stokes parameters $S_0$, $S_1$, $S_2$, $S_3$ of the input radiation were characterized by measuring the transmitted light intensity for certain combinations of $(\alpha, \beta)$ as
\begin{equation}
S_0= I(0^\circ,0^\circ) + I(90^\circ,90^\circ)
\end{equation}
\begin{equation}
S_1= I(0^\circ,0^\circ) - I(90^\circ,90^\circ)
\end{equation}
\begin{equation}
S_2= I(45^\circ,45^\circ) - I(135^\circ,135^\circ)
\end{equation}
\begin{equation}
S_3= \frac{I(45^\circ,0^\circ) - I(135^\circ,0^\circ) - S_2 \cos \delta}{\sin \delta}
\end{equation}
and the degree of polarization (DoP) was calculated by

\begin{equation}
DoP = \frac{\sqrt{{S_1}^2+{S_2}^2+{S_3}^2}}{S_0}
\end{equation}

We note that the equation for $S_3$ is general and can be applied to characterize  $S_3$ using a waveplate with arbitrary phase retardation $\delta$. When a perfect quarter-waveplate $\delta = 90^{\circ}$ is adopted, the equation reduces to the conventional form $S_3 = I(45^{\circ}, 0^{\circ}) - I(135^{\circ}, 0^{\circ})$.

All the signals for the Stokes parameters measurement were collected at a sample temperature of T=493.2K. A quarter-waveplate rated at 7 $\mu m$ was used in the SPARTES system. Since we were interested in the polarimetric properties in a relatively broad spectral range (5.5 - 7.7$\mu m$), the wavelength-dependence of the phase retardation $\delta$ was taken into account, and the general form of the equation was applied to evaluate $S_3$ accurately. 

To obtain the wavelength-dependent retardation $\delta_\lambda$, the waveplate was placed between two parallel polarizers. The transmitted light spectra were collected when rotating the waveplate to different angles $\theta$. Using the Jones-matrix formalism, the transmittance $T_{\lambda}$ can be expressed as
\begin{equation}
T_{\lambda}(\theta) =  {T_p}^2( 1-\sin ^2 \frac{\delta_\lambda}{2} \sin ^2 2 \theta)
\end{equation}
where $T_p$ is the transmittance of the linear polarizers (along the transmission axis). Then $\delta_\lambda$ $\in(0^\circ,180^\circ)$ was obtained at each wavelength $\lambda$ by
\begin{equation}
\cos (\delta_{\lambda} / 2)=\sqrt{\frac{T_{\lambda}(45^\circ)}{T_{\lambda}(0^\circ)}}
\end{equation}

The measured wavelength-dependent retardation $\delta_\lambda$ of the waveplate is shown in Supplementary Materials Figure~S6. The waveplate has quarter-wave retardation around 7$\mu m$, and a linear dispersion of retardance can be seen in the spectral region of 5.5 - 7.7$\mu m$.

Moreover, the signal intensity of left-handed thermal radiation $S_{LCP}$ and right-handed thermal radiation $S_{RCP}$ were characterized based on the relations that $S_0 = S_{LCP} + S_{RCP}$ and $S_3 = S_{LCP} - S_{RCP}$, i.e.
\begin{equation}
S_{LCP} = \frac{1}{2} (S_0 + S_3)
\end{equation}
\begin{equation}
S_{RCP} = \frac{1}{2} (S_0 - S_3)
\end{equation}

\subsection*{System calibration}

To calibrate both the spectral responsivity and the background noise (environmental radiation) and extract thermal emissivity, a blackbody source (IR-518/301 Blackbody, Infrared Systems) with a known thermal radiation spectrum was used. Instead of using a single-temperature measurement, we implemented a more accurate multi-temperature calibration. To elaborate, the collected total signal in the experiment included the emission from the sample, the environmental thermal radiation, as well as the dark counts of the detector itself. Assuming a sample $\alpha$, at a temperature $T$ and angle $\theta$, the collected total signal can be expressed by
\begin{equation}
S_{\bar{\nu}}^{\alpha T \theta} = Z_{\bar{\nu}} \big[e_{\bar{\nu}}^{\alpha \theta} B_{\bar{\nu}}^{T} + (1 - e_{\bar{\nu}}^{\alpha \theta}) E_{\bar{\nu}} \big] + \eta_{\bar{\nu}}
\end{equation}
where $\bar{\nu}$ denotes the wavenumber; $Z_{\bar{\nu}}$ is the responsivity of our detector; $e_{\bar{\nu}}^{\alpha \theta}$ is the temperature-independent spectral emissivity of the sample; $B_{\bar{\nu}}^{T}$ is the blackbody radiation spectrum; $E_{\bar{\nu}}$ is the environment radiation; and $\eta_{\bar{\nu}}$ is the dark counts of the detector. 

In the calibration, we used a blackbody source ($e_{\bar{\nu}}^{\alpha \theta} = 1$). Thus,
\begin{equation}
S_{\bar{\nu}}^{T} = Z_{\bar{\nu}}B_{\bar{\nu}}^{T} + \eta_{\bar{\nu}}
\end{equation}
we were only left with 2 unknown factors $Z_{\bar{\nu}}$ and $\eta_{\bar{\nu}}$. We measured the signal spectrum $S_{\bar{\nu}}$ at a series of temperature T and made a linear-least-square fitting of $S_{\bar{\nu}}^{T}$ as a function of $B_{\bar{\nu}}^{T}$ at each wavenumber, where the slope of the fitting was $Z_{\bar{\nu}}$, and the y-intercept was $\eta_{\bar{\nu}}$, as
\begin{equation}
Z_{\bar{\nu}} = \frac{dS_{\bar{\nu}}^{T}} {dB_{\bar{\nu}}^{T}}
\end{equation}
\begin{equation}
\eta_{\bar{\nu}} = S_{\bar{\nu}}^T - Z_{\bar{\nu}} B_{\bar{\nu}}^{T}
\end{equation}

After the calibration, the emissivity profile for any unknown sample was obtained by taking the differential thermal emission spectra $S_{\bar{\nu}}^{\alpha T_1 \theta} - S_{\bar{\nu}}^{\alpha T_2 \theta}$ at two different temperatures $T_1$ and $T_2$, as
\begin{equation}
e_{\bar{\nu}}^{\alpha \theta} = \frac{S_{\bar{\nu}}^{\alpha T_1 \theta} - S_{\bar{\nu}}^{\alpha T_2 \theta}} {Z_{\bar{\nu}}(B_{\bar{\nu}}^{T_1} - B_{\bar{\nu}}^{T_2})}
\end{equation}

We note that this multi-temperature calibration approach requires the emissivity $e_{\bar{\nu}}^{\alpha \theta}$ to be temperature-independent. To show this temperature independence and validate our methods, raw thermal radiation signals at different temperatures are provided in Supplementary Materials Figure~S8 and Figure~S9.

\bibliography{scibib}

\bibliographystyle{SciAdv}

\clearpage

\section*{Acknowledgments}
The authors thank F. Kalhor, R. Starko-Bowes and C. Khandekar for useful discussions. The authors thank A. Jishi for experimental assistance. \textbf{Funding}: This work was supported by the Defense Advanced Research Projects Agency (DARPA) Nascent Light-Matter Interactions (NLM) program.  \textbf{Author contributions}: X. W., T. S. and S. R. performed the device fabrication and experimental characterization. X. W., S. B. and Z. J. contributed to the theoretical aspects of the work. X. W., Y. W., L. Q. and D. J. performed the numerical simulations.  X. W. and Z. J. wrote the manuscript. Z. J. supervised the research. \textbf{Competing interests}: The authors declare that they have no competing interests. \textbf{Data and materials availability}: All data needed to evaluate the conclusions in the paper are present in the paper and/or the Supplementary Materials.

\section*{Supplementary Materials}

\begin{itemize}
\item {Figure~S1 to S19}
\item {Supplementary Text}
\end{itemize}
\clearpage

\begin{figure}
    \centering
    \includegraphics[width=1\columnwidth]{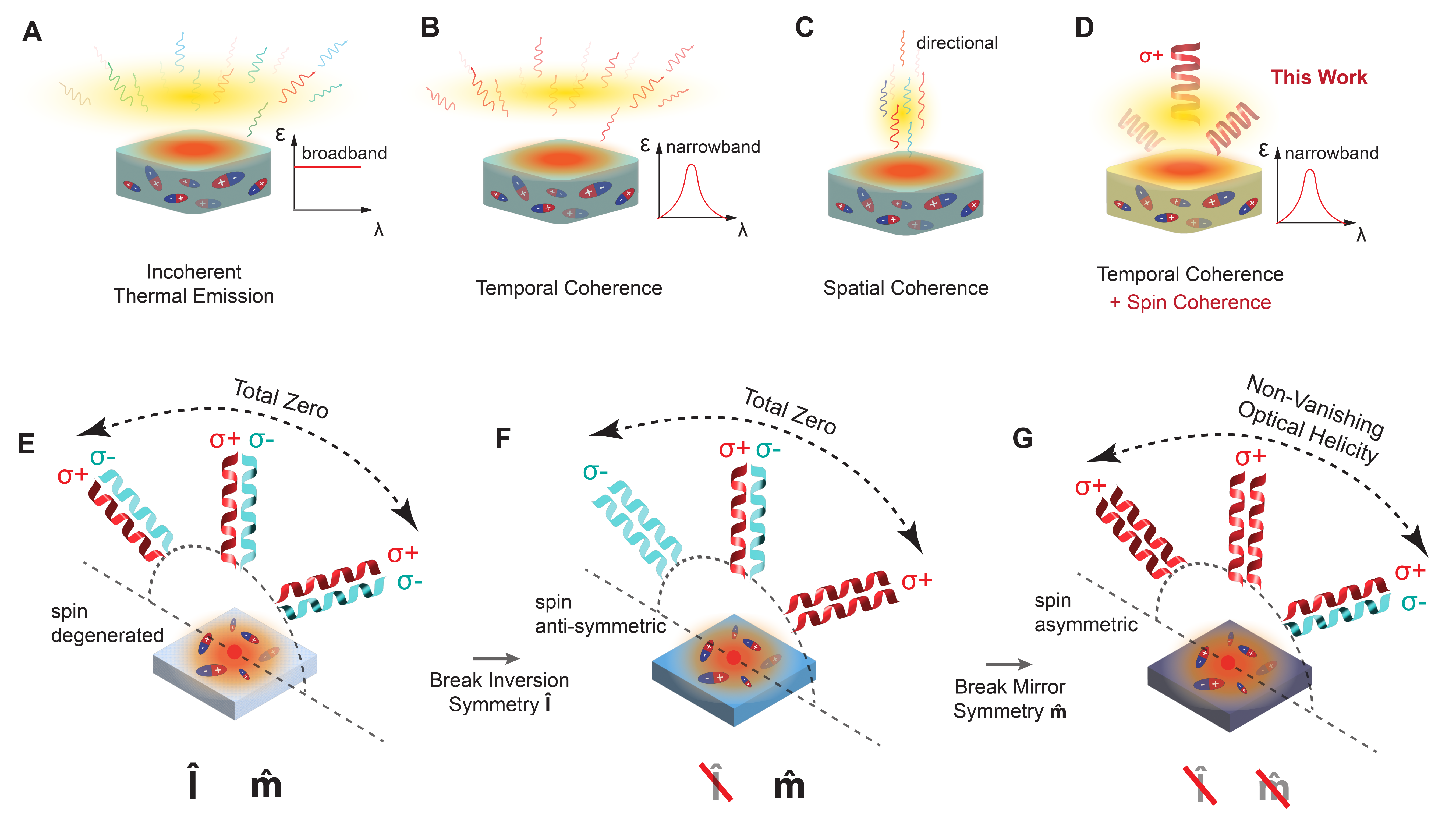}
    \caption{\textbf{Symmetry-based optical spin control of thermal radiation.} \textbf{(A)} Thermal radiation originates from fluctuating dipoles and is thus considered an incoherent signal. It is naturally broadband, omnidirectional and carries no spin angular momentum. \textbf{(B} to \textbf{C)} Recent research efforts aim to impart temporal (B) and spatial (C) coherence in thermal radiation, where narrow-band and directional thermal radiation are demonstrated. \textbf{(D)} In this work, by imparting spin coherence, we achieve effective tailoring of thermal emission in its spectral and spin properties.  \textbf{(E)} Schematic demonstrates that the photon spin characteristics are governed by the symmetries in the 2D system. When both inversion- (\textbf{\^{i}}) and mirror- (\textbf{\^{m}}) symmetries are preserved, the photon spin of thermally radiated photons is degenerate in energy-momentum space. Thus the spin and helicity both vanish. \textbf{(F)} When inversion symmetry is broken, spinning thermal radiation arises at oblique angles. However, the anti-symmetric spin pattern guarantees the spin degeneracy at surface normal and a total-zero optical helicity. \textbf{(G)} In this work, we show photon spin arises in an asymmetric pattern when both inversion- and mirror- symmetries are broken and the non-vanishing optical helicity is observed. }
    \label{fig1}
\end{figure}

\clearpage
\begin{figure}
\centering
\includegraphics[width=1\columnwidth]{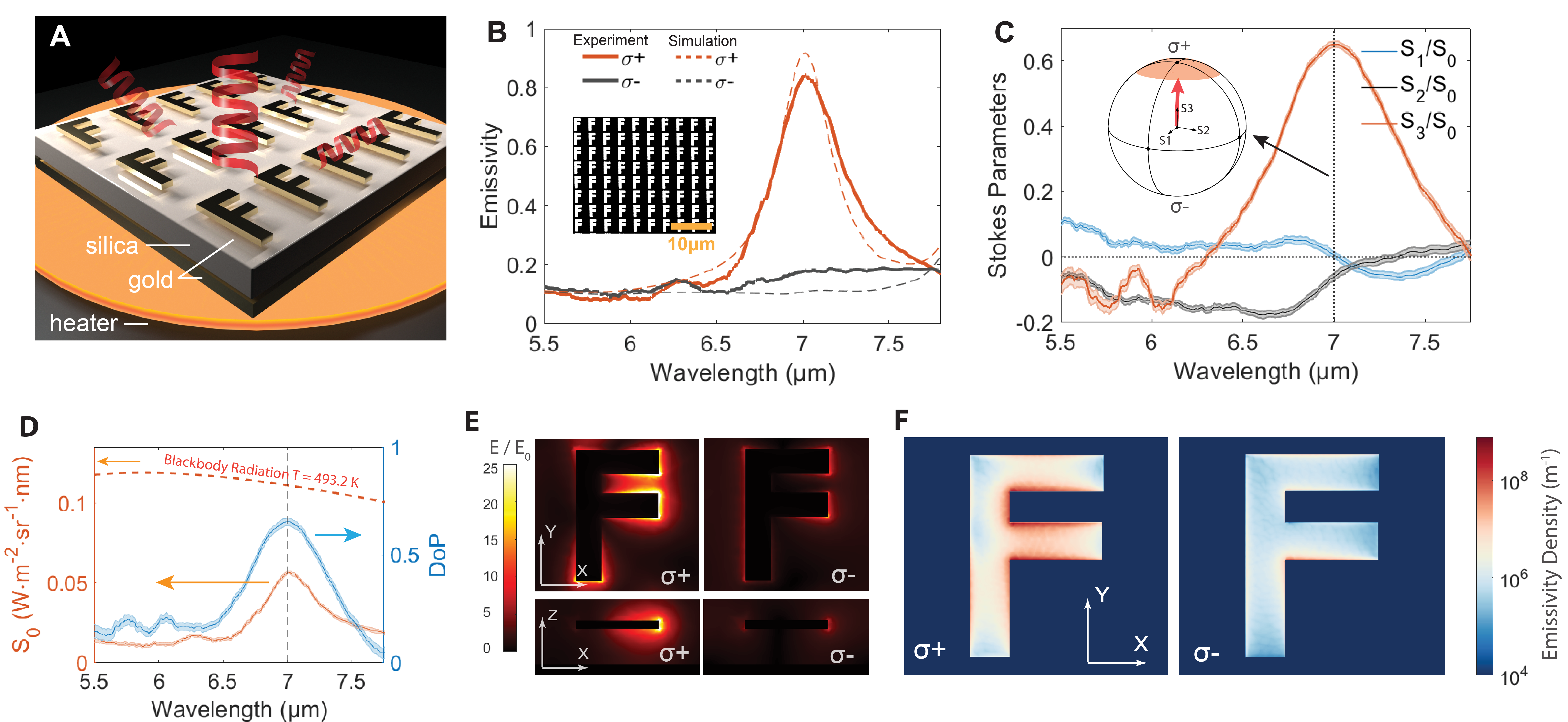}
\caption{\textbf{Spinning thermal radiation and microscopic mechanisms.} \textbf{(A)} Schematic of the designed metasurface with F-shape meta-atoms, where both the inversion- and mirror- symmetries are broken in the pseudo-2D system. \textbf{(B)} The measured LCP ($\sigma+$) and RCP ($\sigma-$) emissivity (solid) of the fabricated devices. They show good agreement with the simulation results (dashed). Inset: SEM imaging of the fabricated device. (Scale bar: 10$\mu$m). The results directly demonstrate the spin-degeneracy is removed and the symmetry-broken metasurface imparts spin coherence in the incoherent thermal fluctuations. \textbf{(C)} Stokes parameters show the full polarization state (shaded areas represent the standard deviations of the measurements). The fabricated device presents a high $S_3$ peak at 7 $\mu$m, while the $S_1$ and $S_2$ are close to zero. Inset: representation of the polarization state in the Poincar\'{e} sphere. \textbf{(D)} $S_0$ and degree of polarization
(DoP)  demonstrate the total intensity and the polarization purity of the thermal radiation signal (shaded areas represent the standard deviations of the measurements). $S_0$ is compared with the blackbody radiation spectrum (dahsed) at the measurement temperature (492.3K). \textbf{(E)} The time-averaged electric field strength (normalized by the field $E_0$ of incident waves) at 7µm under LCP (left) and RCP (right) excitations. The fields are plotted along the XY (top) and XZ (bottom) planes, respectively. \textbf{(F)} Local emissivity density of LCP (left) and RCP radiation (right) in the meta-atom. An evident enhancement is observed for LCP radiation. Figure f indicates the spinning thermal radiation arises from the intrinsic local chirality of the metasurface.}
\label{fig2}
\end{figure}

\clearpage

\begin{figure}
\centering
\includegraphics[width=1\columnwidth]{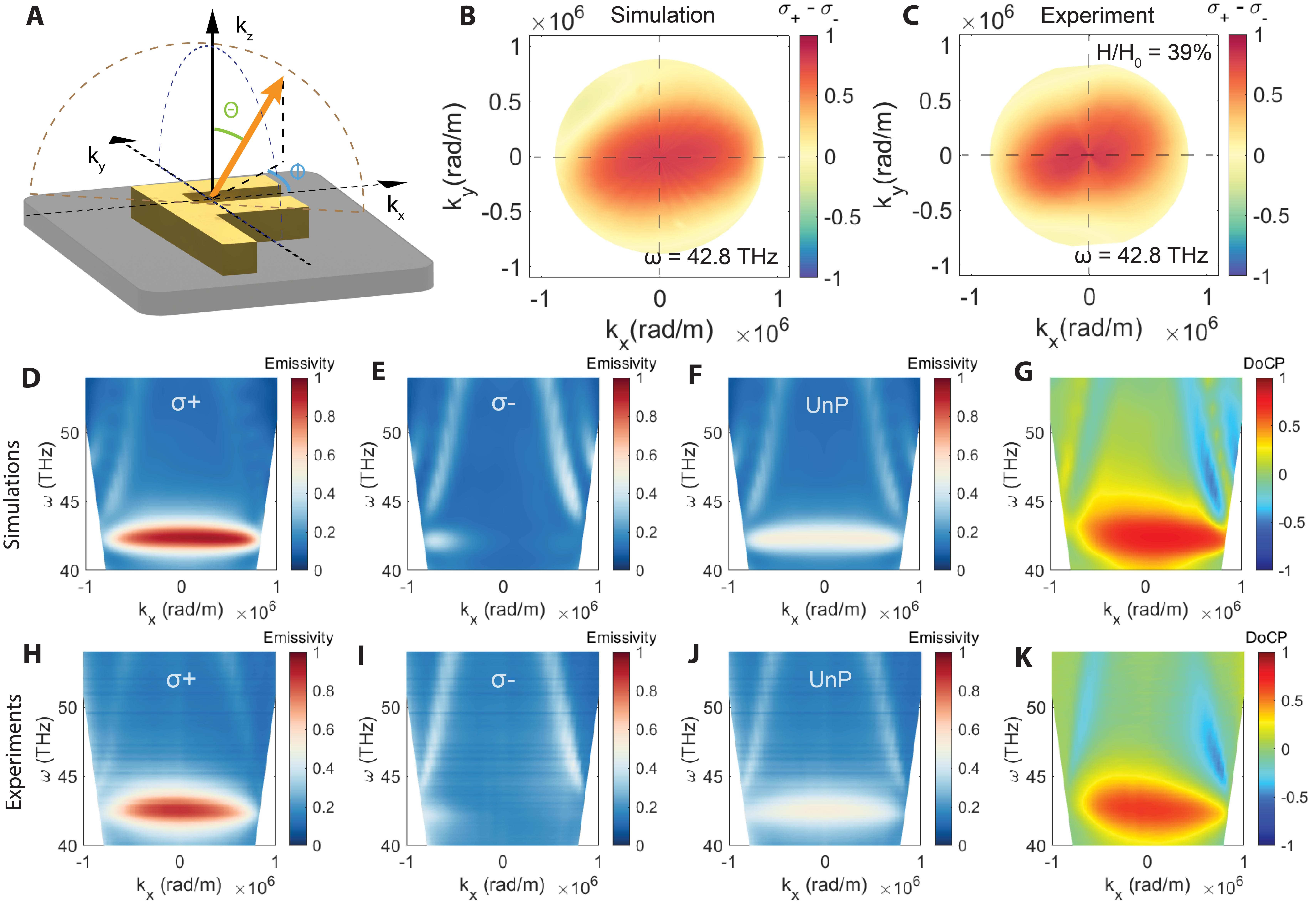}
\caption{\textbf{Spin-polarized angle-resolved thermal emission spectroscopy (SPARTES)}. \textbf{(A)} Schematic of the coordinate system. SPARTES collects thermal radiation signals at different deflection ($\theta$) and azimuth angles ($\phi$) over the far-field hemisphere. \textbf{(B} to \textbf{C)} Calculated (B) and experimentally measured (C) differential emissivity at 7 microns. The thermal radiation is predominantly LCP over the entire hemisphere. The total optical helicity, which is proportional to the integral of the differential emissivity in the k-space, is non-zero. It reaches more than one-third of the fundamental limits ($H/H_0$=39\%).  \textbf{(D} to \textbf{K)} Simulated (top) and measured (bottom) angle-resolved thermal radiation spectra in the energy-momentum space ($k_y=0$) show excellent agreements. The LCP emissivity (D, H), RCP emissivity (E, I), unpolarized emissivity (F, J), and DoCP (G, K) are plotted for comparison. A unique spin-polarized dispersionless band and spin asymmetry are obtained in our fabricated symmetry-broken metasurface. }
\label{fig3} 
\end{figure}

\clearpage

\begin{figure}
\centering
\includegraphics[width=0.75\columnwidth]{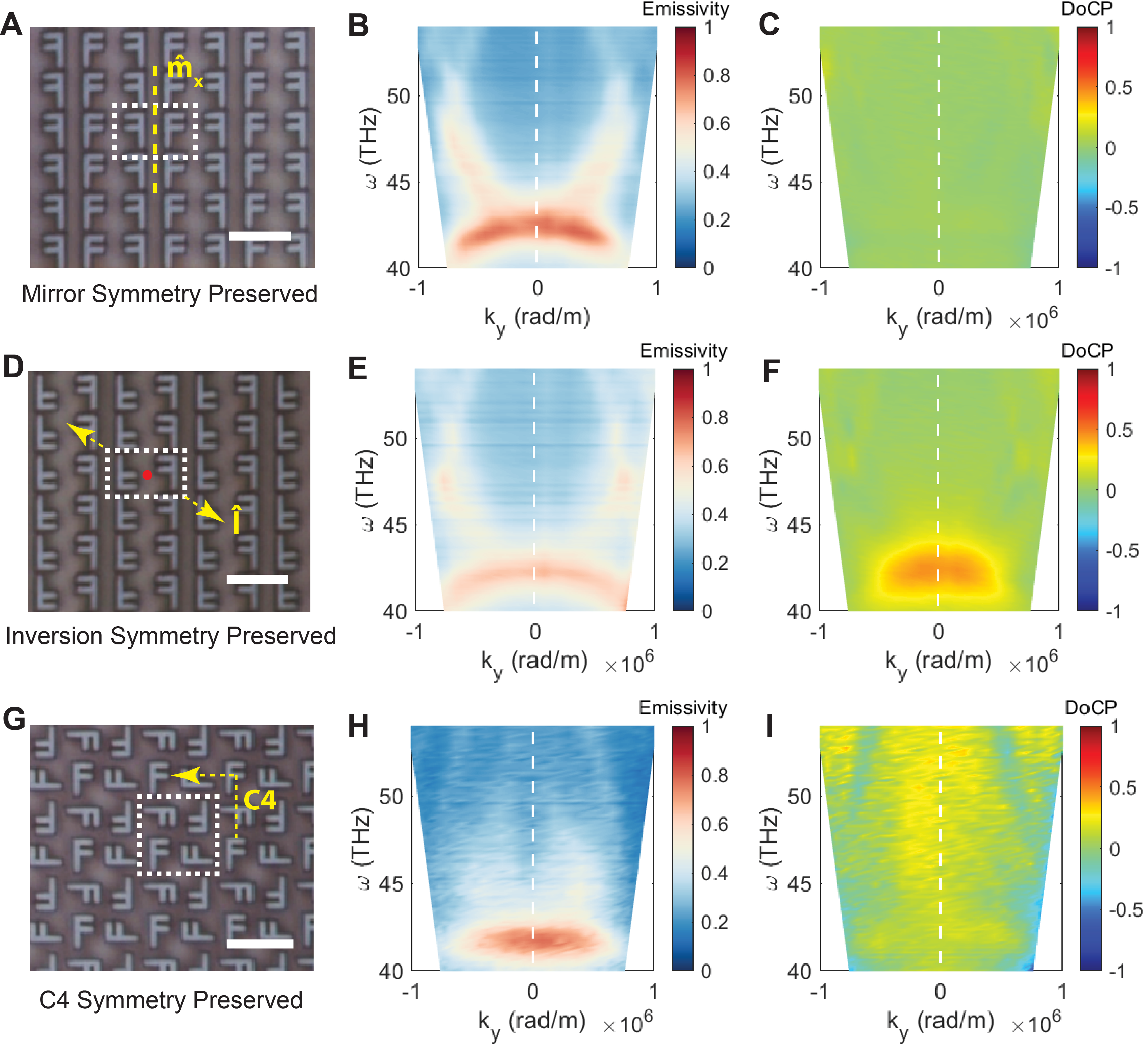}
\caption{\textbf{Symmetry-broken metasurfaces for thermal radiation engineering.} \textbf{(A} to \textbf{I)}. The optical images (left, scale bar: 5 $\mu$m), averaged emissivity spectra (middle), and DoCP (right) are plotted for devices with mirror symmetry (A to C), inversion symmetry (D to F), and four-fold rotational symmetry (G to I), respectively. The average emissivity is symmetric along k=0 in all three cases, which is a manifestation of the reciprocity. The DoCP is zero for the mirror-symmetric device, and symmetric for the inversion- and C4- devices. The results demonstrate that the intertwined spectral and spin properties of thermal radiation can be effectively tailored through symmetry engineering.}
\label{fig4} 
\end{figure}

\end{document}